\documentclass[12pt,preprint]{aastex}

\slugcomment{Submitted to the Astrophysical Journal}

\shorttitle{Altair's Oblateness and Rotation Velocity}

\shortauthors{van Belle et al.}

\begin{document}

%% LaTeX will automatically break titles if they run longer than
%% one line. However, you may use \\ to force a line break if
%% you desire.

\title{Altair's Oblateness and Rotation Velocity from Long-Baseline Interferometry}

%% Use \author, \affil, and the \and command to format
%% author and affiliation information.
%% Note that \email has replaced the old \authoremail command
%% from AASTeX v4.0. You can use \email to mark an email address
%% anywhere in the paper, not just in the front matter.
%% As in the title, you can use \\ to force line breaks.

\author{Gerard T. van Belle\altaffilmark{1}}
\affil{Jet Propulsion Laboratory, California Institute of
Technology, Pasadena, CA 91109\\
gerard@huey.jpl.nasa.gov}

\author{David R. Ciardi}
\affil{Department of Astronomy, University of Florida,
Gainesville, FL, 32611\\
ciardi@astro.ufl.edu}

\author{Robert R. Thompson}
\affil{Jet Propulsion Laboratory, California Institute of Technology, Pasadena, CA 91109, and\\
Department of Physics \& Astronomy, University of Wyoming, Laramie, Wyoming 82071\\
thompson@huey.jpl.nasa.gov}

\author{Rachel L. Akeson}
\affil{Infrared Processing and Analysis Center, California Institute of Technology, Pasadena, CA 91125\\
akeson@huey.jpl.nasa.gov}

\and

\author{Elizabeth A. Lada}
\affil{Department of Astronomy, University of Florida, Gainesville,
FL, 32611\\
lada@astro.ufl.edu}

%% Notice that each of these authors has alternate affiliations, which
%% are identified by the \altaffilmark after each name.  Specify alternate
%% affiliation information with \altaffiltext, with one command per each
%% affiliation.

\altaffiltext{1}{For preprints, please contact: gerard@huey.jpl.nasa.gov.}

%% Mark off your abstract in the ``abstract'' environment. In the manuscript
%% style, abstract will output a Received/Accepted line after the
%% title and affiliation information. No date will appear since the author
%% does not have this information. The dates will be filled in by the
%% editorial office after submission.

\begin{abstract}

We present infrared interferometric angular size measurements for the A7IV-V star
Altair which indicate a non-circular projected disk brightness distribution.
Given the known rapid rotation of this star, we model the data
as arising from an elongated rigid atmosphere.
To first order, an ellipse may be fit to our
interferometric diameter measurements, with major
and minor axes of $2a=3.461\pm0.038$ milliarcseconds (mas) and
$2b=3.037\pm0.069$ mas, respectively, for a difference of
$424\pm79$ microarcseconds ($\mu$as) between $2a$ and $2b$, and with an
axial ratio of $a/b =1.140\pm 0.029$.
Assuming that the apparent oblateness of the photosphere is due to
the star's rapid rotation,
a more rigorous evaluation of the observation data in the context
of a rigidly rotating Roche model shows that an
estimate of $v \sin i = 210\pm 13$ km s$^{-1}$ can be derived
that is independent of spectroscopic techniques. Also derived are
values for the mean effective temperature,
the mean linear radius,
and an observational constraint upon the
the relationship between rotation velocity and stellar inclination
is established.
%Our data also weakly indicates that the most favorable
%inclination of the star is at $i=70-75\deg$, indicating it is
%rotating at approximately half of critical velocity.
Altair is the first main sequence star for which direct observations of
an oblate photosphere have been reported, and the first star for
which $v \sin i$ has been established from observations of the
star's photospheric geometry.

\end{abstract}

%% Keywords should appear after the \end{abstract} command. The uncommented
%% example has been keyed in ApJ style. See the instructions to authors
%% for the journal to which you are submitting your paper to determine
%% what keyword punctuation is appropriate.
\keywords{stars: individual: Altair, infrared: stars, stars:
fundamental parameters, techniques:interferometric}

%% From the front matter, we move on to the body of the paper.
%% In the first two sections, notice the use of the natbib \citep
%% and \citet commands to identify citations.  The citations are
%% tied to the reference list via symbolic KEYs. The KEY corresponds
%% to the KEY in the \bibitem in the reference list below. We have
%% chosen the first three characters of the first author's name plus
%% the last two numeral of the year of publication as our KEY for
%% each reference.

\section{Introduction}

The star Altair ($\alpha$ Aql, HR 7557, HD 187642) is a well-studied
object, being the 12th brightest star in the sky and one of the 50 nearest
stars to the Sun (Allen 1973, Perryman et al. 1997).
It is an A7IV-V main sequence
star (Johnson \& Morgan 1953) and is known to be a rapid rotator,
with an atmosphere that has been extensively
modeled (eg. Gouttebroze et al. 1999).  The measurements of the star's apparent
rotational velocity ($v \sin i$)
range from 190 km s$^{-1}$ (Carpenter et al. 1984)
up to 250 km s$^{-1}$ (Stoeckley 1968),
depending upon the spectral lines used
in the investigation.  These values of $v \sin i$ are a substantial fraction of the
star's estimated critical velocity of 430 km s$^{-1}$
(Gray 1976), where centripetal acceleration
at the stellar equator exceeds gravitational acceleration.

Stellar rotation has been measured observationally for almost a century,
beginning with Schlesinger (1909,1911).
Models of rotating stars have explored the impact of rotation upon both stellar
effective
temperature (Slettebak 1949) and stellar shape (Collins 1963, 1965; Collins
\& Harrington 1966).
Recently models have begun to incorporate the effects of
differential rotation as a function of stellar latitude (Zahn 1992).
Rotation impacts important observable parameters such as
photometry (Collins \& Smith 1985) and
surface brightness distributions, as originally shown by von Zeipel (1924a, 1924b), and rotation has
non-trivial implications upon stellar evolution, as explored
in the various papers by, among others, Claret and Maeder
(cf. Martin \& Claret 1996, Claret 2000,
Maeder 1997, 2000).

Up until now, however, virtually all observational
evidence underpinning the theoretical models has been based
upon velocities inferred from
spectroscopic
line broadening.  While this
technique is both well understood and well developed, it is susceptible
to confusion with other influences upon spectral line widths, such as
various turbulence mechanisms and latitude dependencies of line emission
(see Carpenter et al. 1984 and references therein). An independent
means by which to determine
the parameters governing the structure of centrifugally-distorted stars
% $v \sin i$
would be welcome.

The $3$ milliarcsecond (mas) angular diameter of
Altair was observed thirty years ago by Hanbury Brown and his colleagues
with the
Intensity Interferometer at Narrabri (Hanbury Brown et al. 1967, 1974).
While the authors comment upon the possibility and observational
implications of this
star being
rotationally flattened in their first paper,
they did not explicitly solve for this
possibility, due to insufficient data to constrain
an oblate model (Davis 2000).  As such, Jordahl (1972) examines the Intensity
Interferometer results in the context of
of apparent disk brightness distribution
resulting from stellar rotation theory, although
this is done from the perspective of
its effects upon the average angular diameter.

Herein we report the determination of the overall diameter and
projected shape of Altair upon the sky from near-infrared,
long-baseline interferometric measurements taken with the Palomar
Testbed Interferometer (PTI). PTI is an 85 \& 110 m H- \& K-band (1.6
$\mu$m \& 2.2 $\mu$m) interferometer located at Palomar
Observatory and is described in detail in Colavita et al.
(1999). PTI has a minimum K-band fringe spacing of $\approx 4.3$ mas
at the sky position of Altair, making this particular object
readily resolvable.

\textit{Direct observation} of the stellar disk
%we may be able to
can
provide unique insight into basic stellar
parameters.
The measured angular size in conjunction with the
bolometric
flux and distance yields constraints on parameters such
as effective temperature and linear radii, both of which remain quantities
poorly established empirically for virtually all stars.
Upon fitting a family of rotating models for the
projected stellar photosphere upon the sky, we further demonstrate that a unique
value for $v \sin i$ may be derived from the interferometric data.

%Also,
%our angular size measurements and derived quantities have
%implications regarding the nature of evolution among rapidly rotating
%stars.

The PTI observations that produced these results are discussed in
section 2, detailing source selection and observation.  In section 3, the
procedures used in establishing the stellar parameters for the
stars observed are discussed: the parameters include
spectral type, bolometric flux, major and minor axial angular sizes, effective
temperature and linear radius.
Finally, in section 4, we
demonstrate that an apparent rotational velocity
and other observational parameters may be derived from
Altair's oblateness by fitting the data
with the appropriate family of Roche models.

\section{Observations}

The interferometric observable used for these measurements is the
fringe contrast or visibility (squared) of an observed brightness
distribution on the sky. Normalized in the interval $[0 : 1]$, a
single star exhibits monochromatic visibility modulus in a uniform
disk model given by
\begin{equation}\label{eqn_UDdisk}
V^2 = {\left[{2J_1(\theta_{UD}\pi B  \lambda^{-1})
\over \theta_{UD}\pi B  \lambda^{-1}}\right]}^2,
\end{equation}
where $J_1$ is the first-order Bessel
function, $B$ is the projected baseline vector magnitude at the
star position, $\theta_{UD}$ is the apparent angular diameter of
the star, and $\lambda$ is the wavelength of the interferometric
observation. The $V^2$ observables used in our Altair study are
the synthetic wideband $V^2$'s, given by an incoherent
signal-to-noise (SNR)
weighted average $V^2$ of the 5 narrowband channels in the PTI
spectrometer (Colavita 1999).  In a similar fashion,
incoherent
SNR-weighted average bandpasses $\lambda$ were determined from the raw data.
The PTI
H and K wavebands are excellent matches to the CIT photometric
system (Colavita et al. 1999; Elias et al. 1982, 1983). Separate
calibrations and fits to the narrowband and synthetic wideband
$V^2$ data sets yield statistically consistent results, with the
synthetic wideband data exhibiting superior SNR.
Consequently, we will present only the results from the synthetic
wideband data.

Altair was observed in conjunction with Vega and objects in our
calibrator list by PTI at 2.2~$\mu$m on 7 nights between 1999 May
25 and 2000 July 27.  For three of the nights, PTI's N-W 85m baseline
was utilized; for the remaining four nights, the N-S 110m baseline was used; the
results from each baseline are consistent across all nights.
Altair, along with calibration objects, was
observed multiple times during each of these nights, and each
observation, or scan, was approximately 130 s long. For each scan
we computed a mean $V^2$-value from the scan data, and the error
in the $V^2$ estimate from the rms internal scatter (Colavita
1999). Altair was always observed in combination with one or two
calibration sources within $3\deg$ on the sky. For our study we
have used two main sequence stars as calibration objects: HD
187691 (F8V) and HD 187923 (G0V). The stars are expected to be
nearly unresolved by the interferometer with predicted angular
sizes less than 0.75 mas; expected
angular size and error were based upon a blackbody radiator angular size
inferred from available broadband photometry, particularly in the
near-infrared (Gezari et al. 1996). Clearly, many stars deviate
significantly from blackbody behavior (cf. van Belle et al.
1999); however, the
%hot stars ($T_{EFF} > 5000$K)
main sequence stars of F and G spectral type
selected as
primary calibrators should not deviate sharply from blackbody
behavior.
These objects were additionally
selected to be slow apparent rotators, with $v \sin i <$ 20 km
s$^{-1}$ (Uesugi \& Fukuda 1982). Table 1 lists the relevant
physical parameters for the calibration objects.

The calibration of Altair $V^2$ data is performed by estimating
the interferometer system visibility ($V_{sys}^2$) using
calibration sources with model angular diameters and then
normalizing the raw Altair visibility by $V_{sys}^2$ to estimate
the $V^2$ measured by an ideal interferometer at that epoch
(Mozurkewich et al. 1991; Boden et al. 1998). Uncertainties in the
system visibility and the calibrated target visibility are
inferred from internal scatter among the data in a scan and
standard error-propagation calculations. More detail
on PTI's target \& calibrator selection, data reduction (van Belle
et al. 1999) and technical aspects (Colavita et al. 1999) is
available in the literature. Calibrating our Altair data set with
respect to the two calibration objects listed in Table 1 results
in a total of 27 calibrated scans on Altair over 7 nights in 1999
and 2000. Calibrating our two calibration objects against each
other produced no evidence of systematics, with both objects
delivering reduced $V^2$'s $\approx$ 1, as expected.
Our calibrated synthetic wideband
$V^2$ measurements are summarized in Table 2, along with derived
values as discussed in section 3.  Altair's $V^2$ measurements are plotted
versus spatial frequency in Figure 1, and the uniform disk angular
size versus projected baseline angle on the sky are plotted for Altair and
Vega in Figure 2.

%% To print this table by itself, uncomment the two lines below, as well
%% as the \end{document} command at the end of the file.
%% Otherwise, use \input to insert the file into another
%% LaTeX document. See the end of the accompanying manuscript sample file,
%% sample.tex, for an example of the \input command.

% \documentclass{aastex}
% \begin{document}

\begin{deluxetable}{cccccl}
\tablecolumns{6}
\tablewidth{0pc}
\tablecaption{Calibration Sources.}
\tablehead{
\colhead{Source} & \colhead{$\theta_{EST}$} & \colhead{Distance from}
& \colhead{Spectral}& \colhead{$v \sin i$}& \colhead{Notes} \\
\colhead{} & \colhead{(mas)} & \colhead{Altair (deg)} &
\colhead{Type} & \colhead{(km s$^{-1}$)}& \colhead{} } \startdata
HD187691 & $0.72 \pm 0.10$ & 1.6 & F8V&5 & Primary calibrator\\
HD187923 & $0.51 \pm 0.10$ & 2.8 & G0V&15 & \\
\enddata
\end{deluxetable}

% \end{document}

%%
%% End of file `table.tex'.

%% To print this table by itself, uncomment the two lines below, as well
%% as the \end{document} command at the end of the file.
%% Otherwise, use \input to insert the file into another
%% LaTeX document. See the end of the accompanying manuscript sample file,
%% sample.tex, for an example of the \input command.

% \documentclass{aastex}
% \begin{document}

\begin{deluxetable}{ccccccc}
\tablecolumns{6}
\tabletypesize{\scriptsize}
\tablewidth{0pc}
\tablecaption{The observed data for Altair.}
\tablehead{
\colhead{}   & \colhead{}    & \colhead{} &
\colhead{}    & \colhead{Projected}   & \colhead{Position} & \colhead{Uniform Disk}\\
\colhead{MJD}   & \colhead{Wavelength\tablenotemark{a}}    & \colhead{Normalized} &
\colhead{Hour Angle}    & \colhead{Baseline}   & \colhead{Angle\tablenotemark{b}} & \colhead{Ang. Size}\\
\colhead{ }   & \colhead{($\mu$m)}    & \colhead{ $V^2$} &
\colhead{(hr)}    & \colhead{(m)}   & \colhead{(deg)}& \colhead{(mas)}
}
\startdata
51323.41 & 2.204 & $0.164 \pm 0.010$ & -1.57 & 106.58 & 237 & $3.368 \pm 0.048$  \\
51323.43 & 2.204 & $0.174 \pm 0.013$ & -1.12 & 104.78 & 240 & $3.379 \pm 0.058$  \\
51323.46 & 2.205 & $0.195 \pm 0.014$ & -0.40 & 101.54 & 245 & $3.390 \pm 0.063$  \\
51323.48 & 2.202 & $0.207 \pm 0.015$ & 0.05 & 99.56 & 249 & $3.400 \pm 0.065$  \\
51729.34 & 2.242 & $0.192 \pm 0.007$ & -0.59 & 102.43 & 243 & $3.428 \pm 0.032$  \\
51729.37 & 2.241 & $0.229 \pm 0.010$ & 0.12 & 99.22 & 249 & $3.374 \pm 0.044$  \\
51729.39 & 2.240 & $0.241 \pm 0.009$ & 0.64 & 97.17 & 254 & $3.389 \pm 0.040$  \\
51730.35 & 2.248 & $0.217 \pm 0.030$ & -0.38 & 101.47 & 245 & $3.360 \pm 0.124$  \\
51730.36 & 2.249 & $0.257 \pm 0.043$ & -0.10 & 100.22 & 247 & $3.233 \pm 0.167$  \\
51731.33 & 2.247 & $0.174 \pm 0.023$ & -0.65 & 102.70 & 243 & $3.512 \pm 0.105$  \\
51731.35 & 2.246 & $0.197 \pm 0.030$ & -0.37 & 101.41 & 245 & $3.449 \pm 0.133$  \\
51731.35 & 2.246 & $0.197 \pm 0.021$ & -0.34 & 101.27 & 245 & $3.451 \pm 0.094$  \\
51731.36 & 2.248 & $0.210 \pm 0.036$ & -0.09 & 100.16 & 247 & $3.434 \pm 0.155$  \\
51749.29 & 2.232 & $0.462 \pm 0.062$ & -0.59 & 86.46 & 195 & $2.875 \pm 0.213$  \\
51749.30 & 2.232 & $0.473 \pm 0.044$ & -0.35 & 86.19 & 196 & $2.851 \pm 0.166$  \\
51749.35 & 2.233 & $0.428 \pm 0.043$ & 0.82 & 80.62 & 200 & $3.229 \pm 0.174$  \\
51749.36 & 2.233 & $0.422 \pm 0.042$ & 1.07 & 78.54 & 201 & $3.339 \pm 0.175$  \\
51751.23 & 2.228 & $0.418 \pm 0.016$ & -1.77 & 83.24 & 193 & $3.163 \pm 0.065$  \\
51751.24 & 2.225 & $0.415 \pm 0.021$ & -1.59 & 84.22 & 193 & $3.131 \pm 0.082$  \\
51751.26 & 2.232 & $0.419 \pm 0.017$ & -1.20 & 85.73 & 194 & $3.070 \pm 0.065$  \\
51751.33 & 2.228 & $0.449 \pm 0.024$ & 0.61 & 82.12 & 199 & $3.081 \pm 0.093$  \\
51751.34 & 2.229 & $0.437 \pm 0.025$ & 0.79 & 80.81 & 199 & $3.181 \pm 0.099$  \\
51752.22 & 2.226 & $0.440 \pm 0.018$ & -2.01 & 81.70 & 193 & $3.130 \pm 0.073$  \\
51752.23 & 2.227 & $0.422 \pm 0.015$ & -1.84 & 82.84 & 193 & $3.160 \pm 0.059$  \\
51752.25 & 2.228 & $0.400 \pm 0.021$ & -1.24 & 85.63 & 194 & $3.142 \pm 0.084$  \\
51752.26 & 2.229 & $0.397 \pm 0.021$ & -1.07 & 86.05 & 194 & $3.143 \pm 0.083$  \\
51752.28 & 2.227 & $0.376 \pm 0.022$ & -0.65 & 86.47 & 195 & $3.208 \pm 0.086$  \\
\enddata
\tablenotetext{a}{SNR-weighted average wavelength of the narrowband channels used
to construct the SNR-weighted average $V^2$}
\tablenotetext{b}{PA is east of north}
\end{deluxetable}

% \end{document}

%% In this section, we use  the \subsection command to set off
%% a subsection.  \footnote is used to insert a footnote to the text.

%% Observe the use of the LaTeX \label
%% command after the \subsection to give a symbolic KEY to the
%% subsection for cross-referencing in a \ref command.
%% You can use LaTeX's \ref and \label commands to keep track of
%% cross-references to sections, equations, tables, and figures.
%% That way, if you change the order of any elements, LaTeX will
%% automatically renumber them.

%% This section also includes several of the displayed math environments
%% mentioned in the Author Guide.

\section{Stellar Parameters}

\subsection{Spectral Type \& Bolometric Flux}

Although varying spectral subtypes and luminosity classes are given for Altair,
its spectral type is generally accepted to be A7IV-V (Johnson \& Morgan 1953,
Gliese \& Jahreiss 1991). Bolometric flux was taken from the calibration of Alonso et
al. (1994), who calculate it to be $F_{BOL}=1217\pm46\times  10^8$ erg
cm$^{-2}$ s$^{-1}$.

\subsection{Apparent Stellar Disk}\label{sec_app_disk}

Once a normalized value for
$V^2$ has been obtained,
the simplest interpretation is to fit a uniform disk angular size as presented
in Equation \ref{eqn_UDdisk}.
Altair, with $\theta_{UD}<4$ mas, falls well within the monotonic region of the
uniform disk
function for the 110 m baseline of PTI at 2.2 $\mu$m.
The normalized values for $V^2$ for
each observation are listed in Table 2, with their associated
observation Julian Date, wavelength, hour angle, projected interferometer
baseline and rotation angle, and uniform disk angular size.

Fitting a single global value of $\theta_{UD}$ to the $V^2$ data ensemble results in a mean
uniform disk size of $3.317\pm0.013$ mas with a reduced chi-squared per
degree of freedom (DOF) of $\chi^2/$DOF$=3.71$.
However, as seen in Figure 1, this fit systematically underestimates $V^2$ near
$37$ M$\lambda$ and overestimates near $45$ M$\lambda$.
The smaller spatial frequencies correspond to PTI's N-W baseline, which
is rotated approximately
50 degrees from PTI's N-S baseline, which
was used to obtain the data at larger spatial frequencies.
This discrepancy
can be addressed by relaxing the assumption of spherical symmetry and
including the position angle of the observations in the fit.
A spherical gaseous
star will deform when rotating; such a
shape projected onto the sky will appear, to first order, as an ellipse.
For given physical situations, the true geometry of a rotating star
will depart from that of an ellipsoid
at the 5-6\% level, and we will return to this
in a much more precise fashion
in section \ref{sec_discussion}.  However, such a fit is useful as
a metric to initially establish the position angle dependence of our
angular size data.
Using the basic equation for an ellipse,
\begin{equation}\label{eqn_ellipse}
\theta_{UD}(\alpha) = {2ab \over \sqrt{a^2\sin^2(\alpha-\alpha_0)+
b^2\cos^2(\alpha-\alpha_0) } }
\end{equation}
we may solve for a projection angle-dependent angular size, where
$2a$, $2b$ are the major and minor axes of the ellipse on the sky
in mas, respectively, and $\alpha_0$ is the orientation angle of
the ellipse on the sky, where $\alpha_0=0$ corresponds to the
minor axis pointing to the N on the sky. Fitting equation
\ref{eqn_ellipse} to the data in Table 2, we find that
$2a=3.403\pm0.031$ mas, $2b=2.986\pm0.066$ mas, and
$\alpha_0=25\pm9\deg$ with $\chi^2/$DOF$=0.53$.  An illustration of
these fits and the data is seen in Figure 3.

In contrast to this finding is contemporaneous data taken of Vega.
Data for both stars and their respective calibrators
were taken within an hour of each other during
observing runs in 1999 and 2000.
Vega is best fit by a $3.223\pm0.008$ mas circular disk, with
$\chi^2/$DOF$=0.45$; no regular deviations in the Vega $V^2$ data are seen similar
to the Altair data.  An ellipsoidal fit to the observed Vega visibilities
results in an axial ratio of $a/b=1.024\pm0.032$ with $\chi^2/$DOF$=0.38$, which is of negligible significance statistically.
The Vega observations are discussed in detail in Ciardi et al. (2001).  The uniform disk
sizes versus baseline projection angle for both Vega and Altair can be seen in
Figure 2.  This figure shows that, in contrast to our Vega data, Altair's $V^2$ data are poorly
fit by a single uniform disk model.
Our surprise at this result was somewhat mitigated by corroborating preliminary
indications from the Navy Prototype Optical Interferometer (NPOI)
that their Altair results also exhibit sky position angle dependencies
(Nordgren 2000).

Other potential causes for Altair's departure from circularly symmetric $V^2$ data may be ruled out.
If Altair were either a true or line-of-sight binary star, our interpretation
of the
$V^2$ variations
with baseline length and position angle
would be incorrect.
However, Altair is indicated to be a single
star from astrometric investigations (Gatewood \& de Jonge 1995, Perryman et al. 1997)
in contrast to earlier
indications to the contrary (Russell et al. 1978).
This result is consistent with the recent HST NICMOS investigation of the star
indicating no nearby late-type or substellar companions (Kuchner \& Brown 2000), who rule out
the possibility of companions with $\Delta J \leq 10$
further than 1.4" from the primary, which corresponds
to 3.3 AU at the distance of Altair.
The possibility of a line-of-sight companion may be further investigated given
the fact that Altair is known to be a star with a large proper motion
of roughly 0.65 arcseconds per
year (Perryman et al. 1997).  Examining the 40-50 year old plates from the first
Palomar Sky Survey indicates that Altair's $\approx30$ arcsec of proper
motion did not move it to within 5 arcsec of any objects brighter
than $V\approx10$ magnitude.  Assuming an average color of $V-K=3$,
we can infer no line-of-sight companions with $K<7$. As such,
the Altair/companion ratio would be $\Delta K > 6$.
As demonstrated in Boden et al. (1998),
PTI is largely insensitive to companions with $\Delta K > 4$.
Thus, the possibility of the $V^2$ variations arising from binarity
is strongly ruled out.

We also consider other potential deviations of the apparent
disk of Altair from that of a uniform brightness distribution.
%Ruling out the possibility of binarity affecting the data,
%there are other potential deviations of the apparent disk of Altair upon
%sky from that of a uniform brightness distribution that must be considered as well.
The presence of limb darkening will
affect a star's observed visibility curve and potentially bias our results.
For a slowly rotating star, this effect is
independent of stellar latitude and is
observed to be an increased dimming of the stellar disk from
center to limb.
For Altair's relatively compact stellar atmosphere, the
general effect of compositional limb darkening
upon the observed visibility curve out to the first null
is negligible at 2.2 $\mu$m.
Linear limb darkening for a non-rotating Altair ($\log g=4.0, T_{EFF}=7750$K; Claret et al. 1995)
is $u(2.2\mu$m$)=0.203$ for the linear limb darkening characterization
$I(\mu) = I(1)(1-u(1-\mu))$,
where $u$ is the linear limb darkening coefficient, $\mu=\cos \theta$ describes the angle
between the line of sight and the emergent flux, and $I(1)$ is the monochromatic
specific intensity at the center of the disk.  Computing a visibility curve from
this center-to-limb
brightness profile and comparing it to that of
a uniform disk indicates that angular sizes for slowly rotating stars
derived under the assumption of a uniform disk fit will be undersized by a
factor of 1.017.  These brightness profiles and their resultant visibility functions are seen in Figure 4.

However, for a rapidly rotating star, this phenomenon takes on an additional
latitude dependence, often referred to in the literature as gravity-darkening
or -brightening (eg. Claret 2000).  As first shown by von Zeipel (1924a),
the polar zones
of stars distorted by rapid rotation will be hotter than their equatorial zones,
because the poles are closer to the center of the star.
The consequential non-uniform flux distribution over the stellar
surface affects a star's visibility curve.  However, as is the case with a
slowly rotating star, the impact is for the visibility curve to depart from that of
a uniform disk primarily
after the first zero, as also seen in Figure 4.  We computed visibility curves for a
center-to-limb brightness profile as above, but with an additional 25\% brightness
at the poles covering 20\% of the star's surface, at a variety of stellar orientations relative
to the interferometer
baseline.
These parameters are consistent with Altair models found in Jordahl (1972), which
for $i=60\deg$ range in temperature from $T=7100$K to $T=9300$; between the
temperature extremums, flux at $2.2\mu$m will increase by a factor of 1.5.
Comparison of the resultant visibility curves indicated that the uniform disk
angular sizes derived for a rapidly rotating star like Altair will deviate from the true
angular sizes by factors of 1.011 to 1.029, depending upon star orientation.

Figure 4 shows the strip brightness distribution and resultant visibility
curve for a uniform disk, a $u=0.203$ limb darkened disk, and a limb darkened disk
with a random bright spot.  A Monte Carlo simulation of the various visibility curves from
random orientations of such a star
indicated that angular sizes derived using uniform disk fits would underestimate
the star's true size by $1.017\pm0.006$.  Similar simulations of limb darkened stars with larger (25\%)
or brighter (50\%) spots show underestimates of $1.020\pm0.013$ and $1.021\pm0.017$, respectively,
once again depending upon the star's orientation upon the sky.  The overall scaling implied by these surface
brightness distributions do not account for the `step'
in $V^2$ data seen in Figure 1, and the scale of the discontinuity in the data is approximately 3.5 times
larger than the limb darkening scaling implied from our strip brightness distribution modeling.

On average, the size underestimate due to limb- or gravity-darkening is a marginal adjustment
and is included in our data reduction process merely for the sake
of completeness; the multiplicative factor of $1.017\pm0.006$ is sufficient to
convert our $\theta_{UD}$ sizes
to Rosseland (photospheric) angular sizes, $\theta_R$.  Our resultant major
and minor axes are
$a_R=3.461\pm0.038$ mas, $b_R=3.037\pm0.069$ mas.  For an equivalent circular area
projected upon the sky, we have $\overline{\theta}_R=3.242\pm0.041$.  The axial ratio
$a_R/b_R$ is $1.140\pm0.029$, and the difference between the axes is
$a_R-b_R=424\pm79 \mu$as.

\subsection{Effective Temperature}\label{sec_teff}

Although we may compute a single effective temperature from our data on Altair,
it must be stressed that this will be nothing more than a mathematical
construct derived from geometrical considerations for the purposes of characterizing
the gross properties of the star.
The stellar effective temperature $T_{EFF}$ is defined in terms
of the star's luminosity and radius by $L = 4\pi \sigma R^2
T_{EFF}^4$. Rewriting this equation in terms of angular diameter
and bolometric flux $F_{BOL}$, a value of $T_{EFF}$ was calculated from the
flux and mean Rosseland diameter $\overline{\theta}_R$ using
\begin{equation}\label{eqn_teff}
T_{EFF} = 2341 \times {\left({F_{BOL} \over
{\overline{\theta}}_R^2}\right)}^{1/4}
= 2341 \times {\left({F_{BOL} \over
a_Rb_R}\right)}^{1/4}
\end{equation}
the units of $F_{BOL}$ are $10^{-8}$ erg/cm$^2$s, and $\overline{\theta}_R$, $a_R$, $b_R$
are in mas. The error in $T_{EFF}$ is calculated from the usual
propagation of errors applied to equation \ref{eqn_teff}.  The resultant mean
$T_{EFF}$ for
Altair is determined here to be $7680 \pm 90$K.  This single value for
effective temperature is solely derived from geometric
considerations and is probably an inadequate true characterization of the
stellar surface.  As mentioned in section 3.2, Altair models that account
for gravity darkening effects (Jordahl 1972) show temperatures that
range with stellar latitude from 7100K to 9300K or more, depending upon
rotation speed.

Previous estimates of Altair's
$T_{EFF}$ range from $8250\pm180$K as determined by intensity interferometry
(Hanbury Brown et al. 1967), to 8080K from modelling
(Malagnini \& Morossi 1990), while
Blackwell et al. (1979) infer 7588K from the infrared flux method (IRFM).  The discrepancy between
our value and Hanbury Brown et al.'s is attributable to two effects:
recent (and presumably more accurate) bolometric flux estimates
and our angular size (and hence, derived
$T_{EFF}$)
is larger.  The
discrepancy in $\overline{\theta}_R$ is most likely due to the effect of either limb darkening, which is
greater (and harder to estimate) at visible wavelengths (as used by the Intensity Interferometer),
or limited sampling.
For values of $T_{EFF}$ derived using just the angular size as indicated from the major
axis $a_R$, the inferred temperatures would be 250K too low; using just $b_R$, 250K too high.

A larger implication of this result is the potential inaccuracy of effective
temperatures derived from angular diameters at single projections across the disks of rotationally
distorted stars.
This effect can be as significant as limb darkening in ascertaining a star's $T_{EFF}$, an effect
which is expected to be routinely considered in all studies of stellar effective temperature.

\subsection{Linear Radius}

If we take the
parallax for Altair, $\pi = 194\pm0.94$ mas, as determined by Hipparcos
(Perryman et al. 1997), and combine it with a mean
angular diameter of $\overline{\theta}_{R} = 3.242\pm0.041$ mas obtained in
section 3.2, we obtain an average photospheric radius $R = 1.794\pm0.023 R_\odot$.
(Following the unfortunately contradictory conventions in the literature, we shall present
angular sizes in terms of diameters, and linear sizes in terms of radii.)
However, as discussed in section \ref{sec_app_disk}, the exact radius of Altair
is not best fit by a single number.
From the apparent major and minor axes, we may quantify the radii
of the star at the extreme latitudes.  At the equator, the stellar
radius should simply
be equal to the projected major axis, $R_a=1.915\pm0.023 R_\odot$.
At the poles, the radius will be less than or equal to the projected
minor axis, $R_b=1.681\pm0.039 R_\odot$.
Given the unknown inclination of the star relative to our line of sight, it
is likely the star's radius at the poles is even smaller.
% We shall discuss the physical implications of this ellipsoidal fit in section
%\ref{sec_discussion}.

As with Altair's effective temperature, the mean linear size we derive is in good agreement
with $R=1.82 R_\odot$ derived from the infrared flux method (Shallis \& Blackwell 1980).  From their
parallax measurement in conjunction with the Hanbury Brown et al. (1967) value
for $\theta_R$,
Gatewood \& de Jonge (1995) estimate $R = 1.63\pm0.08 R_\odot$.
A summary of Altair's stellar parameters is presented in Table 3.

\section{Discussion\label{sec_discussion}}

The key to understanding the peculiar diameter results for Altair
lies in its rapid rotation.  Different values for $v \sin i$ for
the star have been derived from spectral line broadening profiles,
depending upon the spectral line used: Carpenter et al. (1984)
derive $190\pm38$ km s$^{-1}$ from $IUE$ UV data; Freire Ferrero
et al. (1978) derive 220 km s$^{-1}$ from the visible Ca II lines;
Stoeckley (1968) derives $250\pm10$ km s$^{-1}$ by observing
visible Mg I and Ca II lines.

In contrast to these values, the observed rotational velocity of
Vega is roughly a factor of ten lower, at $<20$ km s$^{-1}$
(Freire Ferrero et al. 1983), which is consistent with its
apparent lack of oblateness. While Vega has been reported to be a
rapid rotator, this inference has  been made in conjunction with
the deduction that it is very nearly pole-on (Gulliver et al.
1994) with $i=5-6 \deg$, an orientation which would present the
star to the interferometer as nearly circular upon the sky. As
discussed in section \ref{sec_app_disk} and in Ciardi et al.
(2001), our data from PTI indicating circular symmetry for the
disk of Vega are consistent with this conclusion.

We may more precisely investigate the rotational distortion of the
photosphere of a star.  The force of centripetal acceleration at
the equator, resulting from the rotation, offsets the effect of
gravitation owing to the mass of the star. Under the conditions of
hydrostatic equilibrium, uniform rotation,
%, accounting for the
%presence of centripetal acceleration at the stellar equator
and a point mass gravitational potential, we may derive the
equatorial rotational velocity, assuming we view the star at an
inclination angle $i$.  As developed in the work by Collins (1963,1965)
and presented in Jordahl (1972), the equation of shape for such a star under
rotation may be written as
\begin{equation}
{GM \over R_p(\omega) } = {GM \over R(\theta,\omega) }+{1 \over 2}
\omega^2 R(\theta,\omega)^2 \sin^2\theta
\end{equation}
or
\begin{equation}\label{eqn5}
1 = {1 \over r(\theta)} + {4 \over 27} u^2 r(\theta,u)^2 \sin^2 \theta
\end{equation}
with substitutions for the normalized radius,
\begin{equation}
r(\theta,\omega) = {R(\theta,\omega) \over R_p(\omega)},
\end{equation}
and the dimensionless rotational speed $u$, as defined by
\begin{equation}
\omega^2 = u^2 {8 \over 27} {GM \over {R_p}^3(\omega)}
\end{equation}
where $R(\theta,\omega)$ is the stellar radius at colatitude
$\theta$ for a star of mass $M$ rotating at angular velocity
$\omega$, $R_p(\omega)$ is the polar radius for that star,
and $G$ is the gravitational constant.
As is appropriate in utilizing their mathematical constructs, we also adopted the
coordinate system that is graphically illustrated in Collins (1965) and Jordahl (1972).
Clearly there are more recent and/or more complicated models than this
simple Roche model, but for the purposes of this analysis,
we shall consider it sufficient.
Solving for the cubic equation \ref{eqn5} trigonometrically, we can arrive at
an expression for the
colatitude-dependent stellar radius at a rotation speed $u$:
\begin{equation}
r(\theta,u) = {3 \over u \sin \theta} \cos \left[ {\cos ^{-1} (-u \sin \theta) + 4 \pi \over 3} \right]
\end{equation}
For
the following computations, we used values of 1.8 $M_\odot$ for
the mass of Altair, and a parallax of $\pi=194.45\pm0.94$ mas. Such a
Roche model is applicable for a rigidly rotating star, which is
consistent with a fully radiative stellar atmosphere. Fortunately,
for the case of A-type main sequence star Altair, this is a
reasonable expectation.  It is worth noting that, in contrast to
our elliptical approximation in section \ref{sec_app_disk} , this
approach exactly solves for the expected shape of the stellar
limb.

To interpret our radius data, we began by constructing models of
Altair based upon rotation $u$ and polar radius $R_p(\omega)$,
which are sufficient to map the entire surface as a function of
stellar colatitude and longitude.  Model surfaces were
constructed for the full star at intervals of $0.5 \deg$ in both
colatitude and longitude across the whole volume. These models were
then mapped onto the sky and matched to the observational data
through the use of two additional angular parameters, inclination $i$ and
on-sky rotational orientation $\alpha$ at steps of
0.1 deg along the entire circumference of the stellar limb.

%The
%more familiar angular term from rotational velocity studies is the
%inclination of the rotation axis of the star relative to the line
%of sight to the observer.  Secondly, we needed to solve for the
%rotation of the projected stellar surface about that line of
%sight, which was denoted simply by $\alpha$.

Thus, for a given set of four randomized free parameters
$\{u,R_p(\omega),i,\alpha\}$, a 300,000
point volume surface was generated, projected upon the sky, fit to
the angular diameter data, and a $\chi^2/$DOF calculated.  A
multi-dimensional optimization code was then utilized to derive
the best $\{u,R_p(\omega),i,\alpha\}$ solution from the random
starting point, a process that took typically 500 iterations.
(Press et al. 1992)  An exhaustive search of the rotating star
parameter space was used to explore the $\chi^2/$DOF space.
Furthermore, a static grid of $\{u,i\}$ values was explored for
optimal $\{R_p(\omega), \alpha\}$ values to ensure that no local
minima were trapping the optimization code.  The grid consisted of
1,000 points spread uniformly over the space enclosed by
$u=[0:1]$, $i=[0:90]$ and was run
multiple times with random $\{R_p(\omega), \alpha\}$ seed values,
to ensure full mapping of the resultant $\{u,i\}$ $\chi^2/$DOF
surface. Trial runs of the $\chi^2/$DOF minimization technique using
artificial data sets from synthetic stars were able to fully
recover the original four parameter characterization for the
original synthetic star.  The model data sets covered a wider range of
position angles, from $5 \deg$ to $175 \deg$ in $5 \deg$ steps,
but with angular size errors comparable to the Altair dataset,
which on average are 2.3\% per measurement. The $\chi^2/$DOF surface
resulting from the Altair dataset is plotted in figure \ref{fig_chisurf},
where $\{R_p(\omega), \alpha\}$ are optimized for minimum $\chi^2/$DOF
for a given pair of ${u,i}$ coordinates.

There is a noticable trough in the $\chi^2$/DOF surface obtained via
this technique, running from $u=0.77$ at $i=90 \deg$
(corresponding to viewing the star equator-on) to $u=1.00$ at
$i=31.9 \deg$.  Any inclination less than $31.9 \deg$ is
physically inconsistent with our observed data.
Unfortunately, there is no global minima that is differentiated from the other best
fits in a statistically significant manner.  Instead, we fit for
the trough in $\{u,i\}$ $\chi^2/$DOF space, mapping the family of
models enclosed by +1 of $\chi^2/$DOF that
describe the dependency of rotation $u$ upon inclination $i$:
\begin{equation}\label{eqn9}
u = 4.961\times10^{-5}(90-i)^2 + 1.116\times10^{-3}(90-i) + 0.762
\end{equation}
From each member of this family of models, derived values for
angular velocity $\omega$, equatorial radius $R_e(\omega)$,
equatorial velocity $v_e$, and apparent rotation velocity $v \sin
i$ may be derived. An interesting aspect of this family of models
is that they present uniform values for $v \sin i$, $R_e(\omega)$
and $\alpha$.  Taking the ensemble of solutions found in the
trough and averaging the result, we find that $v \sin i = 210 \pm
13$ km/s. The preferred values for $R_e(\omega)$ and $\alpha$ are
$1.8868\pm0.0066$ $R_\odot$ and $-21.6\pm6.2 \deg$ east of north
for the pole, respectively, which are consistent with the values
derived in section \ref{sec_app_disk} for the ellipsoidal
approximation.  A illustration of a $\{u=0.82, i=70 \deg\}$
potential solution for our data is given in figure 6. Examining
predicted equatorial radii from the family of models,
we are able to establish a critical rotation velocity of
\begin{equation}
v_{CRIT} = \omega_c R_e(\omega_c) = \sqrt{GM \over R_e(\omega_c)^3} R_e(\omega_c)= 426
\pm 12 \textrm{km/s},
\end{equation}
given our constant equatorial radius
and assuming a 10\% error on Altair's mass.  This value is consistent with
the value of 430 km/s from Gray (1976).

An improvement to this technique would be to incorporate
latitude dependent limb- and gravity-darkening, derived from a latitude-dependent
temperature profile and projected onto the sky, resulting in a
relationship between baseline projection angle and the limb
darkening parameter $\mu$.  Such an effort, however, is beyond the
scope of this initial investigation into the observational
appearance of rotational distorted photospheres.  A variant of
this approach has already been employed in the visible by Jordahl
(1972), using the average angular size data of Hanbury Brown et al.
(1974). In the near-infrared, as illustrated in section \ref{sec_app_disk}, we
expect this to be a much smaller effect than in the visible, at
the $\approx1\%$ level, which is
at a level much smaller than our typical angular size errors.
A potential next step in the development of this technique would be to
combine multi-wavelength observations, potentially from multiple interferometers,
to fit a ensemble of stellar
models that features latitude-dependent temperatures.

Our derived value for $v \sin i$ is in good agreement with
spectroscopically determined values, as presented in section 3.5.
The dominant source of error in our technique is primarily located
in the angular size data, and secondarily in the mass
estimate. The linear sizes are well constrained by the Hipparcos
parallax, which has only a 0.4\% quoted error, and dominated by
the angular size error estimates of $\approx 2.3\%$.

Conversely, we may take a measured value for $v \sin i$ in conjunction with our
values for stellar radius and attempt to
infer the mass of the star.  Unfortunately, this approach is unduly
sensitive to errors in both the rotational velocity and measures of
the stellar shape.
For example, if we were to take $v \sin i$ for Altair to be 220 km sec$^{-1}$ with a
10\% error (cf. Freire Ferrero et al. 1995 and references therein), we measure
the mass of this A7IV-V star to be $M = 1.74 \pm 0.49 M_\odot$.
Substantial improvement in this
measurement will be challenging: 1\% values for polar, equatorial
radii and $v \sin i$ still return only an 11\% mass.  A 1\% mass from this
technique requires 0.1\% values for radii and rotational velocity, which appears
currently highly challenging for
both interferometry and spectroscopy.

%% To print this table by itself, uncomment the two lines below, as well
%% as the \end{document} command at the end of the file.
%% Otherwise, use \input to insert the file into another
%% LaTeX document. See the end of the accompanying manuscript sample file,
%% sample.tex, for an example of the \input command.

% \documentclass{aastex}
% \begin{document}

\begin{deluxetable}{llcll}
\tablecolumns{5}
\tablewidth{0pc}
\tablecaption{Basic parameters derived from the data and assembled from the literature.}
\tablehead{
&\colhead{Parameter} & \colhead{Value} & \colhead{Units}    & \colhead{Reference}}
\startdata
\multicolumn{2}{l}{Given}\\
&Spectral Type & A7IV-V & & Johnson \& Morgan 1953\\
&Parallax & $194.45 \pm 0.94 $ & mas & Perryman et al. 1997\\
&$\theta$ (0.5 $\mu$m) & $2.98 \pm 0.14 $ & mas & Hanbury Brown et al. 1974 \\
&$F_{BOL}$ & $1217 \pm 46 $ & $10^8$ ${\textrm{erg} \over \textrm{cm}^2 \textrm{s}^1}$ & Alonso et al. 1994 \\
&$M$ & $1.80 $ & $M_\odot$ & Malagnini \& Morossi 1990\\
\multicolumn{2}{l}{Ellipsoidal fit}\\
&$a_R$ & $3.461 \pm 0.038 $ & mas & This work \\
&$b_R$ & $3.037 \pm 0.069 $ & mas & This work \\
&Position angle & $-25 \pm 9 $ & deg & This work \\
&$R_a$ & $1.915 \pm 0.023 $ & $R_\odot$ & This work \\
&$R_b$ & $1.681 \pm 0.039 $ & $R_\odot$ & This work \\
\multicolumn{2}{l}{Roche models}\\
&$R_e(\omega)$ & $1.8868 \pm 0.0066$ & $R_\odot$ & This work\\
&$\alpha$ & $-21.6\pm 6.2$ & $\deg$ & This work\\
&$v \sin i$ & $210  \pm 13 $ & km/s & This work \\
\multicolumn{2}{l}{Additional derived values}\\
&$a_R/b_R$ & $1.140 \pm 0.029 $ &  & This work \\
&$a_R-b_R$ & $424 \pm 79 $ &$\mu$as  & This work \\
&$\overline{\theta}_R$ & $3.242 \pm 0.041 $ & mas & This work \\
&$\overline{R}$ & $1.794 \pm 0.023 $ & $R_\odot$ & This work \\
&$v_{CRIT}$ & $426 \pm 12$ & km/s & This work
\enddata
\end{deluxetable}

%\end{document}

%%
%% End of file `table.tex'.

%Altair represents the first main sequence star for which asymmetry
%observations have been made but it is not the first star to show
%such geometry.

Our measurements represent the first ever direct
asymmetry observations for a main sequence star.
We emphasize here that all other stars with
reported asymmetries are either Mira variables or supergiant stars
(cf. Lattanzi et al. 1997, Monnier et al. 1999). Previous work
indicates that the radii of the Mira and supergiant stars are
hundreds of $R_\odot$ (Haniff, Scholz \& Tuthill 1995, van Belle
et al. 1996, van Belle et al. 1999); equation 4
indicates a carbon star such as V Hya with $v \sin i \approx 14$
(Barnbaum et al. 1995) and $R\approx400R_\odot$ (van Belle et al.
1997) should exhibit an oblateness of $a/b\approx1.11$.  While
this is remarkable in itself, it is perhaps an unsurprising
expectation for these stars with extended atmospheres; in contrast
to that is our finding of asymmetry for the relatively compact
atmosphere of Altair. We are not aware of any other observations
of luminosity class V or IV-V stars that indicate convincing
departure from circularly-symmetric brightness distributions.

Recent studies of Altair's polarization (Piirola 1977, Tinbergen 1982) have indicated no
statistically significant degree of polarization; as such, there is no comparison to be
made between the
apparent axes of the stellar photosphere and the polarization orientation.

\section{Conclusions}

We have measured the apparent oblateness of Altair's disk upon the sky
and inferred its rotational velocity.
This interferometric
measurement of $v \sin i$ is independent
of spectroscopic and photometric means that have characterized all previous
rotational velocity measurement techniques.  Furthermore, we have demonstrated a
technique that, with sufficient data sampling around a stellar limb, has
the potential to recover the inclination of rapidly rotating stars.

A simple examination of the rotational velocity catalog collated by Bernacca \&
Perinotto (1973) indicates there are over 70 known bright ($V<$4)
main sequence stars in the
northern hemisphere that are rapid rotators with $v \sin i >$ 200 km s$^{-1}$;
examination of bright ($V<$8) evolved objects in
de Medeiros \& Mayor (1999) that have $v \sin i >$ 15 km s$^{-1}$ indicates
there are over 70
potential targets as well.  Objects that fit these criteria
should exhibit apparent flattening of their disks at the $\approx 10\%$ level.
Clearly there are plenty of opportunities
to implement this technique with the upcoming generation of
long-baseline optical and infrared interferometers
such as CHARA, NPOI, Keck, and VLTI, which all have multiple baselines
allowing the required stellar disk projection measurements to be
made in much shorter observing times.  Our PTI follow-up observing campaign
of $\alpha$ Leo and other rapidly rotating stars already has initial results
that support this promising line of research.

For almost a century, the three basic methods for
measuring axial rotation of stars have been line
profile analysis, photometric modulation of starlight
due to dark or bright areas on a rotating star,
and radial velocity curve distortions in eclipsing binary systems (Slettebak 1985); of these, line profile
analysis has been the most widely used.
To these three methods, we add this fourth interferometric approach, which
may have particular utility in servicing stars out of the reach of the other three methods.
As noted in Gray (1976), spectroscopic
determination of $v \sin i$ for rapidly rotating main sequence stars
from spectral lines is
made difficult by line broadening; similarly, spectroscopic
determination of $v \sin i$ for
highly evolved objects
is non-trival due to density of features in their
spectra (cf. Kahane et al. 1988, Barnbaum et al. 1995).
Not only is the value we have derived above
in agreement with the velocities derived from various spectral lines, it is
an independent check of $v \sin i$ values derived from spectroscopic or photometric means for
rotationally flattened stars.

\acknowledgments

We would like to thank Mel Dyck for first suggesting to us the possibility
of utilizing oblateness measurements to derive rotational velocity, and gratefully
acknowledge our referee, George W. Collins, II, for challenging
us to extend our result with thoughtful and positive comments.
We acknowledge fruitful discussions with Andy Boden, Michelle
Creech-Eakman, Steve Howell, Michael Scholz and Francis Wilkin.
We would also like to thank the gracious Palomar Mountain staff and
PTI Collaboration for their support of this observing program.
Funding for PTI was provided to the Jet Propulsion Laboratory under its
TOPS (Towards Other Planetary Systems), ASEPS (Astronomical Studies
of Extrasolar Planetary Systems), and Origins programs and from
the JPL Director's Discretionary Fund.
Portions of this work were
performed at the Jet Propulsion Laboratory, California Institute
of Technology under contract with the National Aeronautics and
Space Administration. DRC acknowledges support from NASA WIRE ADP NAG5-6751.
EAL acknowledges support from a Research Corporation Innovation Award and
a Presidential Early Career
Award for Scientists and Engineers to the University of Florida.

%% The reference list follows the main body and any appendices.
%% Use LaTeX's thebibliography environment to mark up your reference list.
%% Note \begin{thebibliography} is followed by an empty set of
%% curly braces.  If you forget this, LaTeX will generate the error
%% "Perhaps a missing \item?".
%%
%% thebibliography produces citations in the text using \bibitem-\cite
%% cross-referencing. Each reference is preceded by a
%% \bibitem command that defines in curly braces the KEY that corresponds
%% to the KEY in the \cite commands (see the first section above).
%% Make sure that you provide a unique KEY for every \bibitem or else the
%% paper will not LaTeX. The square brackets should contain
%% the citation text that LaTeX will insert in
%% place of the \cite commands.

%% We have used macros to produce journal name abbreviations.
%% AASTeX provides a number of these for the more frequently-cited journals.
%% See the Author Guide for a list of them.

%% Note that the style of the \bibitem labels (in []) is slightly
%% different from previous examples.  The natbib system solves a host
%% of citation expression problems, but it is necessary to clearly
%% delimit the year from the author name used in the citation.
%% See the natbib documentation for more details and options.

\begin{figure}\label{fig_altair_v2}
     \epsscale{1.0}
     \plotone{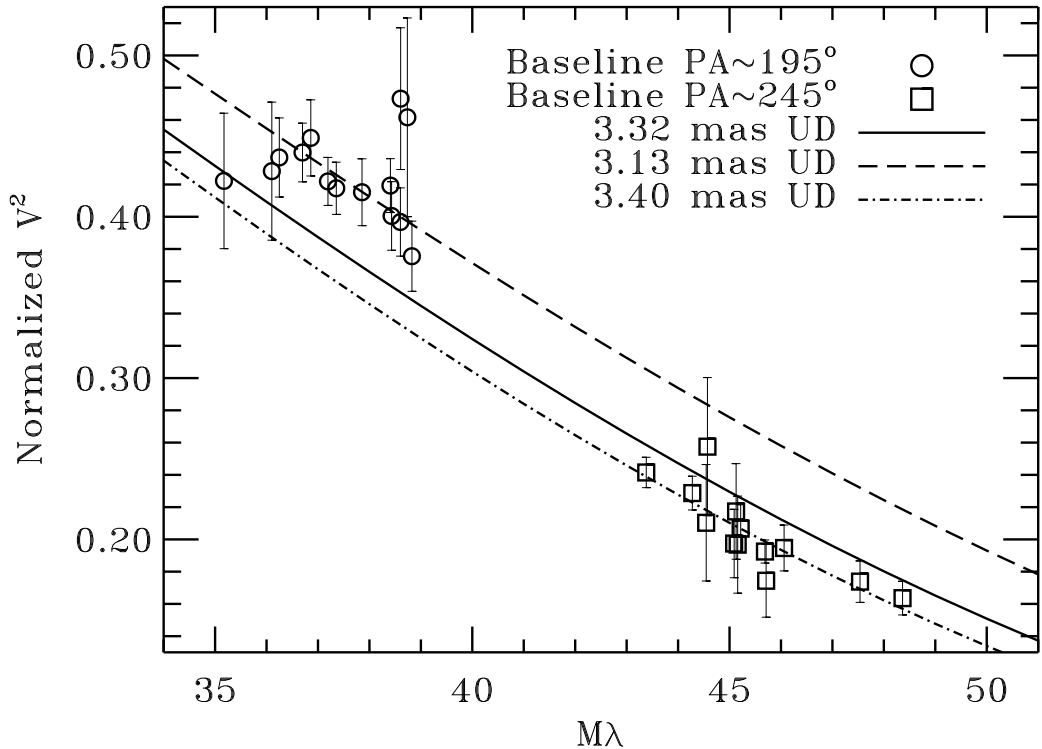}
     \caption{Visibility data for Altair.  The visibility points at $\approx 37$ M$\lambda$ correspond to a
     baseline projection angle of $\approx195\deg$ with a $3.137\pm0.025$ mas uniform disk angular size,
     and the visibility points at $\approx 45$ M$\lambda$ correspond to a
     baseline projection angle of $\approx245\deg$ with a $3.400\pm0.018$ mas angular size.  These
     two baseline projections result from the North-South and North-West baselines of PTI. A single
     3.22 mas disk fit to all of the points clearly is inadequate in fully characterizing the data.}
\end{figure}

\begin{figure}\label{fig_vega_altair}
     \epsscale{0.75}
     \plotone{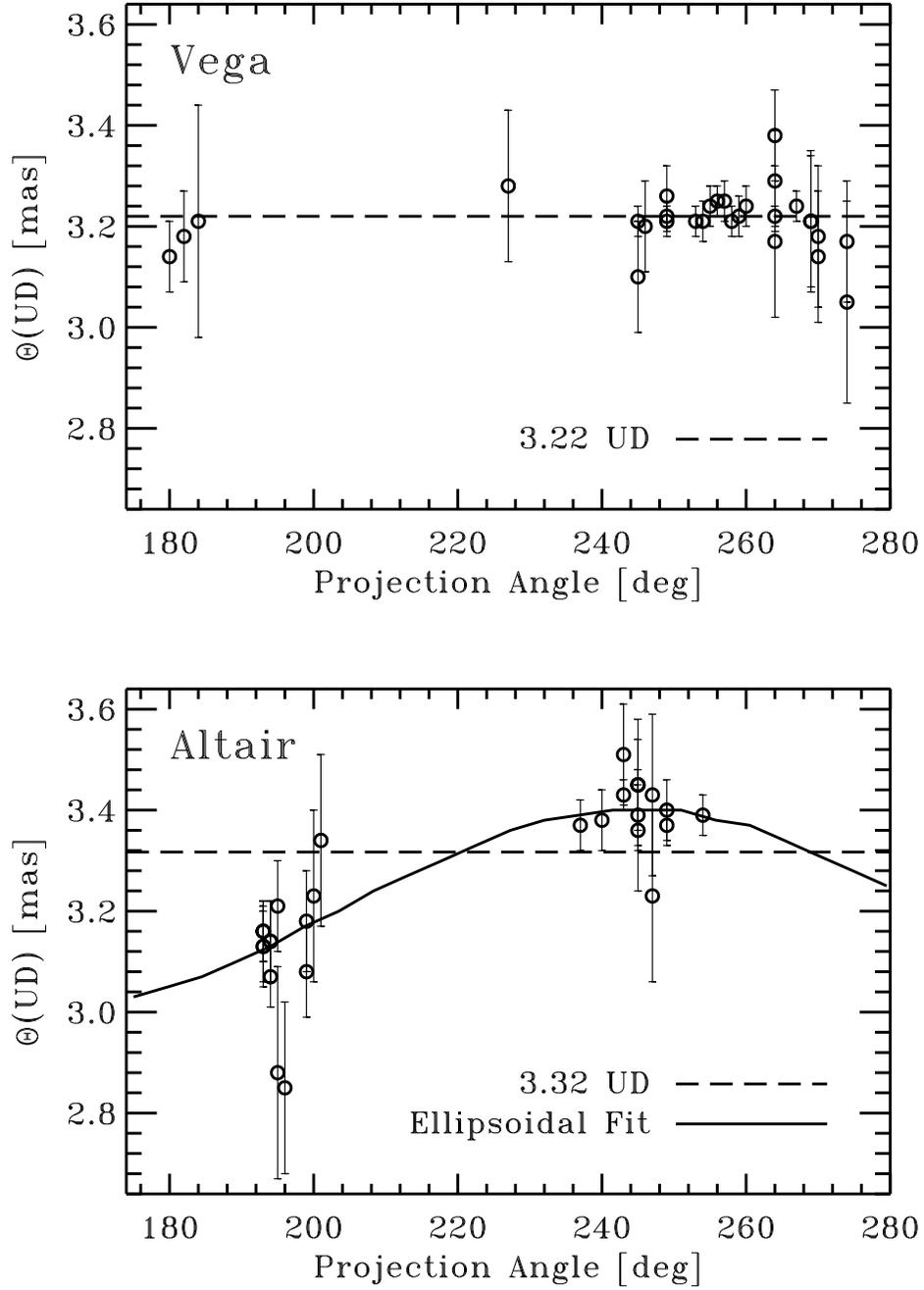}
     \caption{Uniform disk sizes as a function of baseline projection angle for Altair and Vega.}
\end{figure}

\begin{figure}\label{fig_altair_model}
     \epsscale{0.6}
     \plotone{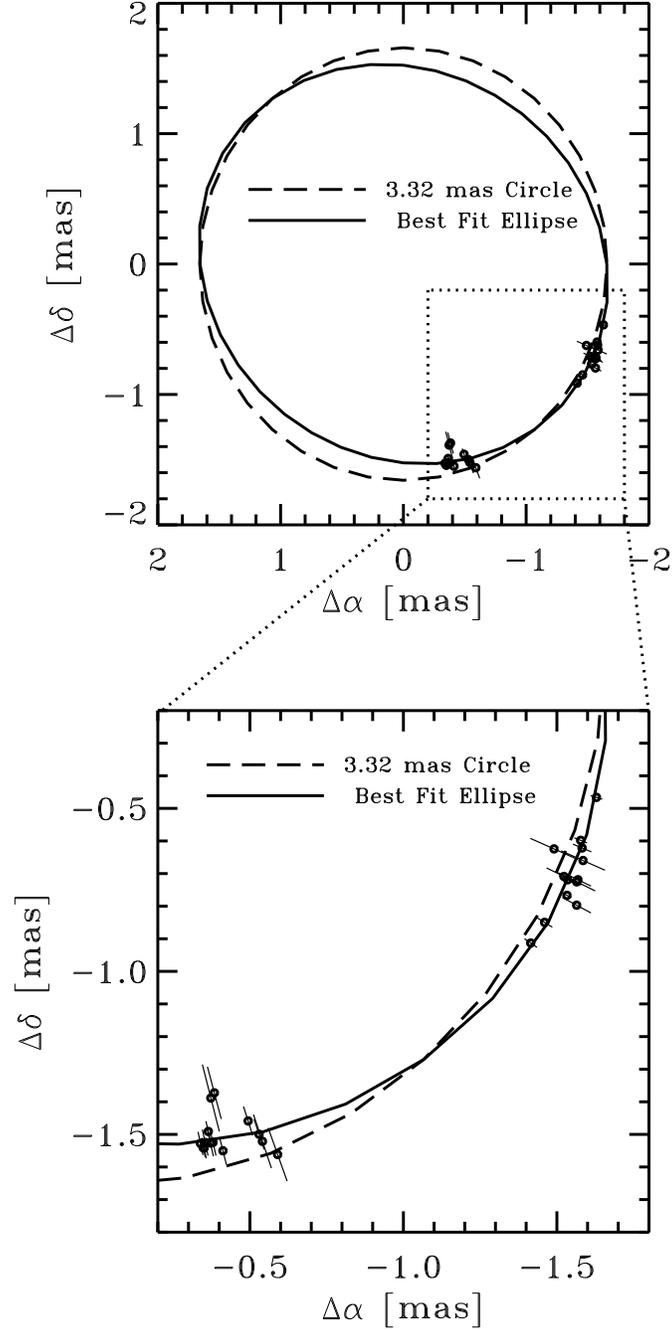}
     \caption{Data points along the limb of Altair.  The data subsets at the two mean position angles each
     contain 13-14 data points. }
\end{figure}

\begin{figure}\label{fig_UDLDdisk}
     \epsscale{0.75}
     \plotone{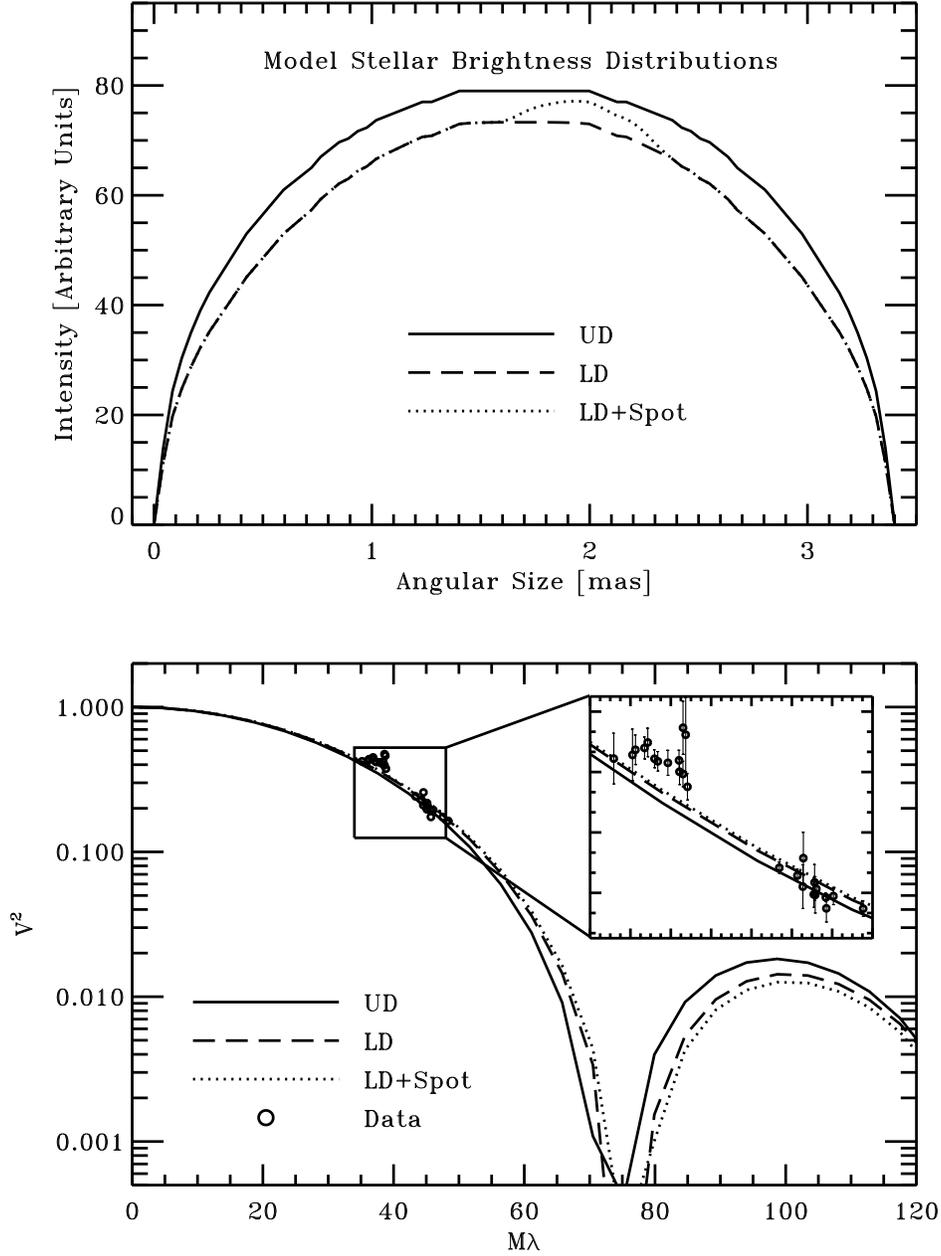}
     \caption{A representative stellar model used in calculating the effect of
both limb darkening and gravity darkening upon the resultant visibility curve.
The upper plot is the 1-D strip brightness for a uniform disk (dotted), limb
darkened disk (solid), and a limb darkened disk with a spot that has a 25\%
brightness enhancement and is 20\% of the stellar disk size (dash).  Below, the
resultant $V^2$ curves are plotted.  A uniform disk fit to the spotted model,
randomly oriented upon the sky, will result in a systematic size estimate that
is a factor of $1.017\pm0.006$ too small. }
\end{figure}

\begin{figure}\label{fig_chisurf}
     \epsscale{0.75}
     \plotone{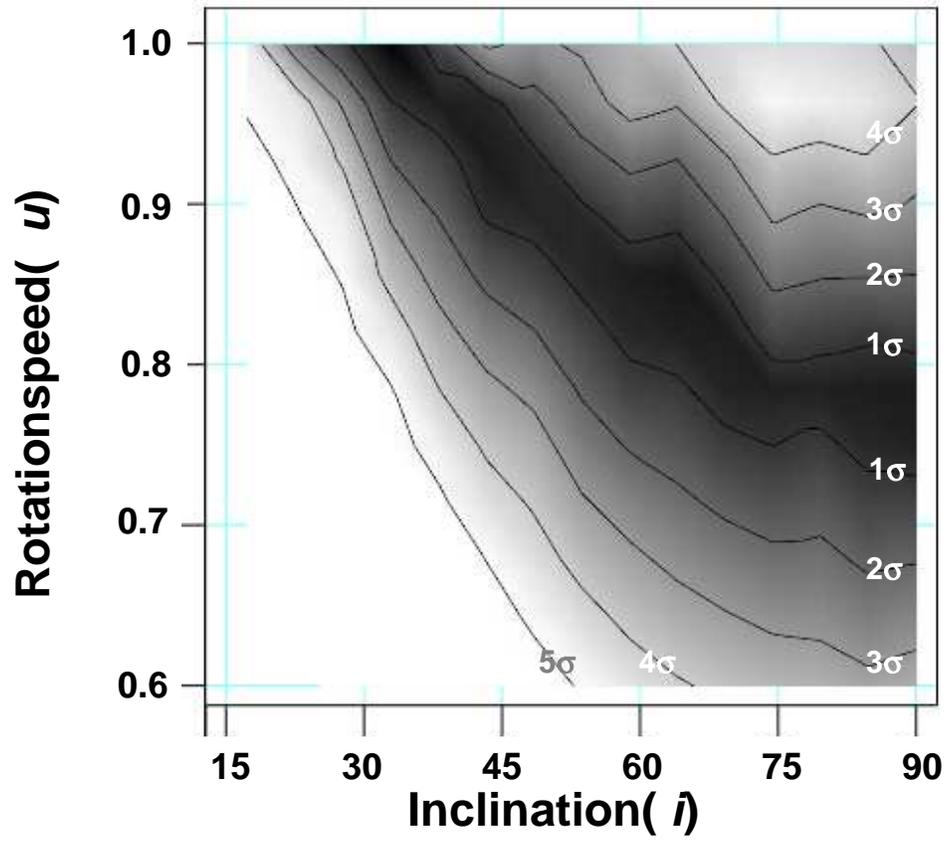}
     \caption{$\chi^2$ / DOF surface for Altair as a function of rotation $u$ and inclination $i$. }
\end{figure}

\begin{figure}\label{fig_am2}
     \epsscale{0.95}
     \plotone{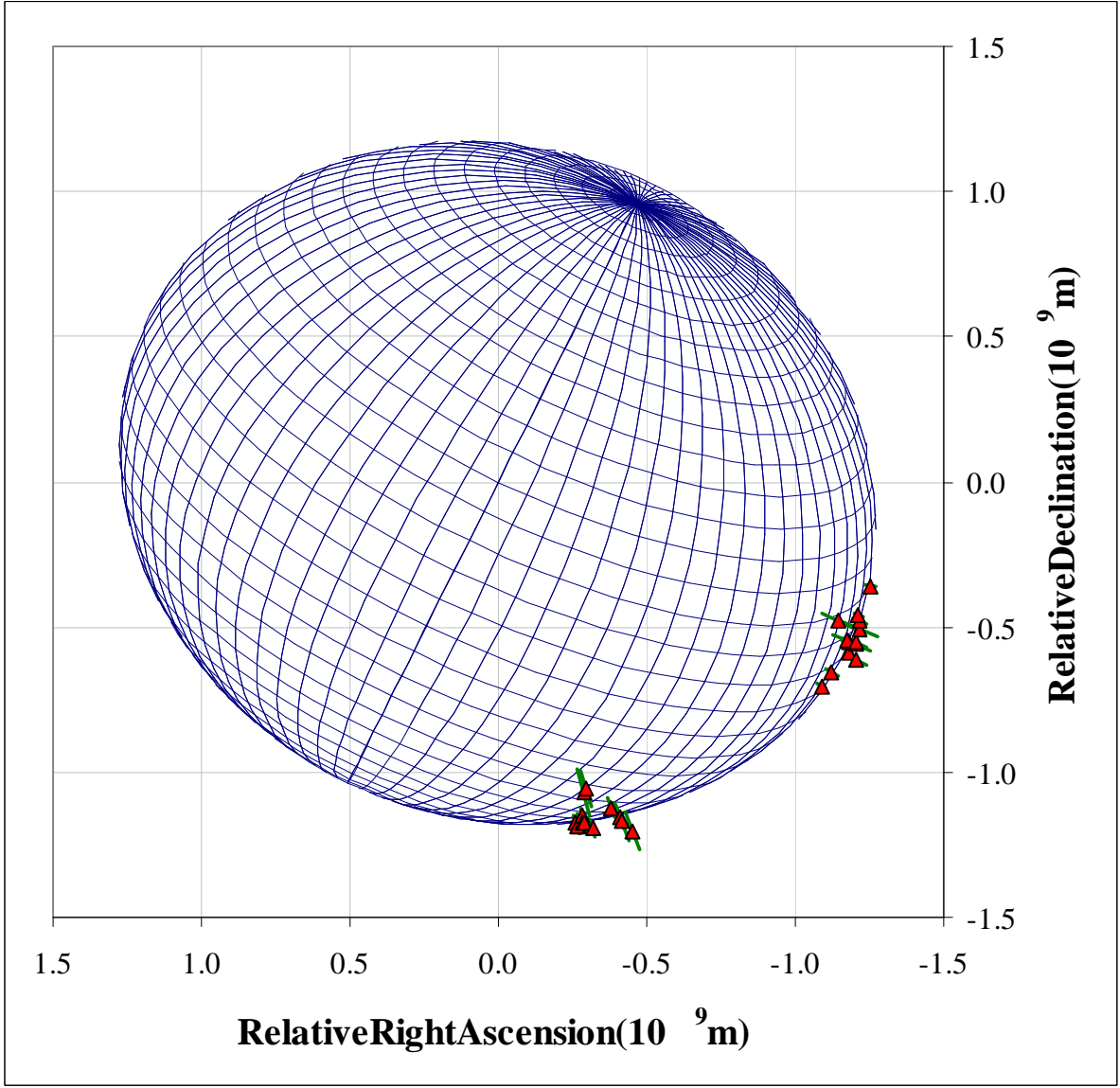}
     \caption{Example 3D model of Altair projected onto the sky ($u=0.82, i=70$), showing the fit of the PTI
     data to the limb of the stellar photosphere.  Units are in meters at the distance
     of Altair.}
\end{figure}

\end{document}